# A Pseudo DNA Cryptography Method

Kang Ning
Email: albertnk@gmail.com

The DNA cryptography is a new and very promising direction in cryptography research. DNA can be used in cryptography for storing and transmitting the information, as well as for computation. Although in its primitive stage, DNA cryptography is shown to be very effective. Currently, several DNA computing algorithms are proposed for quite some cryptography, cryptanalysis and steganography problems, and they are very powerful in these areas. However, the use of the DNA as a means of cryptography has high tech lab requirements and computational limitations, as well as the labor intensive extrapolation means so far. These make the efficient use of DNA cryptography difficult in the security world now. Therefore, more theoretical analysis should be performed before its real applications.

In this project, We do not intended to utilize real DNA to perform the cryptography process; rather, We will introduce a new cryptography method based on central dogma of molecular biology. Since this method simulates some critical processes in central dogma, it is a pseudo DNA cryptography method. The theoretical analysis and experiments show this method to be efficient in computation, storage and transmission; and it is very powerful against certain attacks. Thus, this method can be of many uses in cryptography, such as an enhancement insecurity and speed to the other cryptography methods. There are also extensions and variations to this method, which have enhanced security, effectiveness and applicability.

## 1. Introduction

As some of the modern cryptography algorithms (such as DES, and more recently, MD5) are broken, the new directions of information security are being sought to protect the data. The concept of using DNA computing in the fields of cryptography and steganography is a possible technology that may bring forward a new hope for powerful, or even unbreakable, algorithms.

It is Adleman, with his pioneering work [Adleman, 1994]; set the stage for the new field of bio-computing research. His main idea was to use actual chemistry to solve problems that are either unsolvable by conventional computers, or require an enormous amount of computation. By the use of DNA computing, the Data Encryption Standard (DES) cryptographic protocol can be broken [Boneh, et. al, 1995]. The one-time pad cryptography with DNA strands, and the research on DNA steganography (hiding messages in DNA), are shown in [Gehani, et. al, 2000].

However, researchers in DNA cryptography are still looking at much more theory than practicality. The constraints of its high tech lab requirements and computational limitations, combined with the labor intensive extrapolation means. Thus prevent DNA computing from being of efficient use in today's security world.

In this project, WEam not intended to use real DNA computing, but just use the principle ideas in central dogma of molecular biology to develop my method. The method only simulates the transcription, splicing, and translation process of the central dogma; thus, it is a pseudo DNA cryptography method. It is shown that for information of length *n*, the time complexity of brute force attack to this method is round $O(2^n)$. Experiments also show that the method is very efficient in computation, storage and transmission.

Although this method is efficient, and it is powerful against certain attacks; the partial information contained in the cipher text makes the method not so strong. Therefore, the current method is only suitable for use as an enhancement for the other cryptography methods. However, We have suggested several extensions and variations, which can be made to enhance this method. These optimizations of the methods (such as the use of multiple rounds) have great potential to have much better performance. Since this is a new cryptography method, only some primitive ideas are incorporated in it, and there are still a lot to be done based on it.

## 2. Literature Survey

DNA computing (molecular computing) utilizes the inherent combinational properties of DNA for massively parallel computation. The idea is that with an appropriate setup and enough DNA, one can potentially solve huge mathematical problems by parallel search. Therefore currently, the DNA computing depends primarily on mass storage of DNA and parallel computation.

Ashish Gehani, Thomas LaBean and John Reif have published a paper [Gehani, et. al, 2000], which puts an argument forward that the high level computational ability and incredibly compact information storage media of DNA computing has the possibility of DNA based cryptography based on one time pads. While current practical applications of cryptographic systems based on one-time pads is limited to the confines of conventional electronic media, they argued that small amount of DNA can suffice for a huge one time pad for use in public key infrastructure (PKI).

[Gehani, et. al, 2000] has explained their one-time-pad DNA-based cryptography scheme. They argued that DNA can suffice for a huge one time pad for use in public key infrastructure (PKI). Injecting DNA cryptography into the common PKI scenario, these researchers argue that they have the ability to follow the same inherent pattern of PKI but using the inherent massively parallel computing properties of DNA bonding to perform the encryption and decryption of the public and private keys. In essence, the encryption algorithm used in the transaction can now be much more complex than that in use by conventional encryption methods.

One of the DNA-based cryptography schemes proposed by [Gehani, et. al, 2000] is to use chip-based micro-array technology, and can be generally expressed below

**Encryption**:

· take one or more input DNA strands (considered to be the plaintext message).

· append to them one or more randomly constructed *"secret key"* strands.

· Resulting *"tagged plaintext"* DNA strands are *hidden* by mixing them within many other additional "distracter" DNA strands which might also be constructed by random assembly.

**Decryption**:

· Given knowledge of the “secret key” strands.

· Resolution of DNA strands can be decrypted by a number of possible known recombinant DNA separation methods:

Plaintext message strands may be separated out by hybridization with the complements of the “secret key” strands might be placed in solid support on magnetic beads or on a prepared surface.

These separation steps may be combined with amplification steps and/or PCR.

The procedure is illustrated in the following figure.

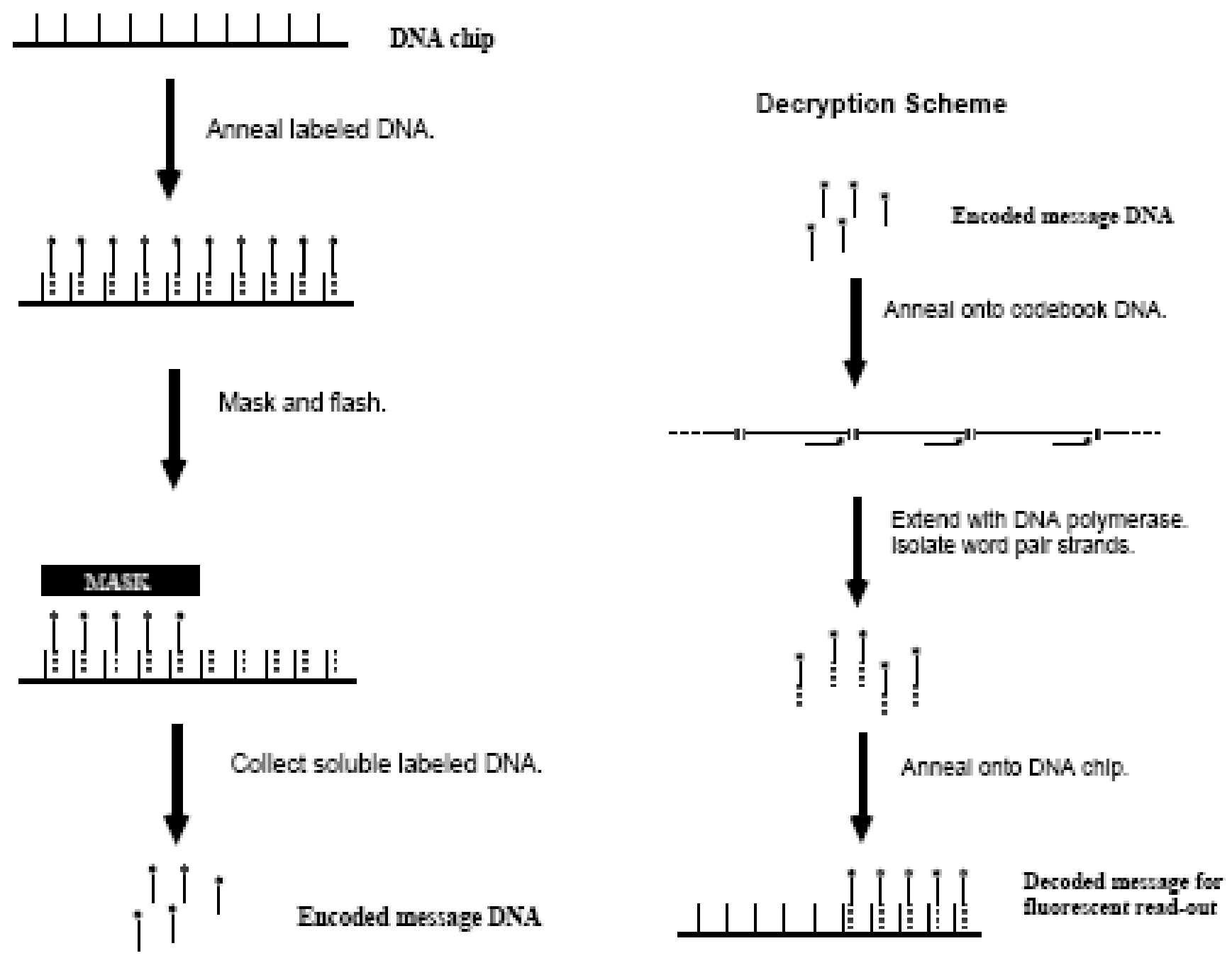

Figure 1: The process of encryption and decryption for DNA-based cryptography.

This process is used in 2D image encryption. They also suggested the use of DNA computing on steganography. The method with improved security is also suggested by these researchers.

It is easy to see that DNA computing is just classical computing, albeit highly parallelized and mass storage. Thus the idea of this form of DNA computing is at great risk in the field of cryptography. Also, such technology is very hard to use outside laboratories currently, both getting the DNA strands, and extracting the results.

## 3. Motivation and Method

The pseudo DNA cryptography method is different from that of the DNA cryptography based on DNA computing. The method does not really use DNA sequences (or oligos), but only the mechanisms of the DNA function; therefore, the method is a kind of pseudo DNA cryptography methods. The cipher /

decipher process of the method is based on the central dogma of the molecular biology, and the process is similar to the DNA transcription, splicing and RNA translation of the real organisms.

### Biological principles

The central dogma of the molecular biology is illustrated in the following figure.

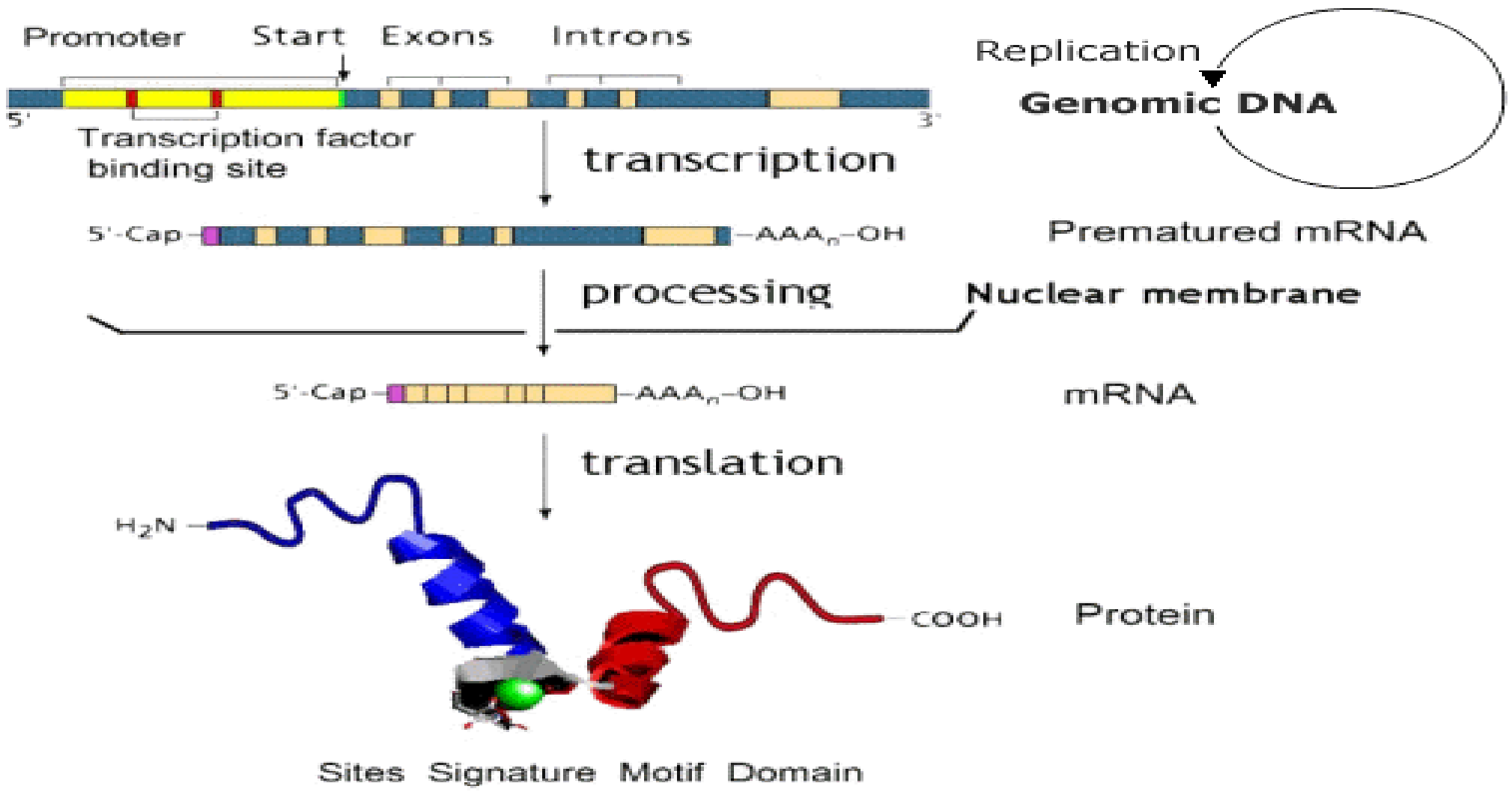

Figure 2: The Central Dogma of molecular biology.

The transcription, splicing and translation processes are briefly explained below

**Transcription and Splicing**: a DNA segment that constitutes a gene is read, starting from the promoter (starting position) of the DNA segment. The non-coding areas (intron) are removed according to certain tags, and the remaing coding areas (extron) are rejoined and caped. Then the sequence is transcribed into a single stranded sequence of mRNA (messenger RNA). The mRNA moves from the nucleus into the cytoplasm.

**Translation**: the mRNA sequence is translated into a sequence of amino acids as the protein is formed. During translation, the ribosome read the fragment starting from certain three-bases, and then the ribosome reads three bases (a codon) at a time from the mRNA and translates them into one amino acid; there are also certain ending three-bases to sign the end of the translation.

Essentially, in the transcription and the splicing steps, introns are cut out, and extrons are kept to form mRNA, which will perform the translation work. In the translation process, codons are translated into the amino acids according to the genetic code table.

### Rationale

The pseudo DNA cryptography method consists of 2 parts, similar to the translation/splicing and transcription processes.

To make the cipher text difficult to decipher, We have made some changes to the original splicing process. Originally, the introns are characterized by their starting and ending codes, which makes the guess about introns relatively easy. We have changed this to another scheme, in which the start codes and the pattern codes specify the introns. The pattern codes are non-continuous patters, which define which parts of the frame to be removed, and which parts to be kept. This makes the guess on introns difficult, since they are now spaced introns. Since the starting and pattern codes can determine the length of the introns, the ending codes are not necessary.

Informally, let Alice be the sender of the information and Bob be the receiver. From Alice's point of view, the information is stored in the binary form, and can be transformed into DNA form (A for 00, C for 01, G for 10, T for 11). Alice also knows the starting codes (codes that indicate the begin of the intron) and pattern codes (codes that define which parts of the frame to be removed, and which parts to be kept) of the introns, so she knows where are the introns in the DNA form of the information and which parts should be removed. Therefore, the 2 parts can be described briefly as below

1. The DNA form of information is scanned by Alice to find out the introns; she records the introns places, and cut out the introns according to the specified pattern. So that the DNA form of data is translated into the mRNA form of data.

2. Alice translates the mRNA form of data into protein form of data according to the genetic code table (61 codons to 20 amino acids).

The protein form of data is then transferred to Bob. The starting and pattern codes of the introns, the places of the introns, the removed spaced introns, and the codon-amino acids mapping of the protein are the keys to decrypt the protein form of data, and they are transferred to Bob through a secure channel (or they are encrypted by Bob's public key, and transferred to Bob).

On Bob's side, he receives keys through the secure channel from Alice (or uses public key protocol to communicate with Alice to receive keys). When he received the protein form of data and the keys, Bob uses the keys to recover mRNA form of data from protein form of data, and then recover DNA form of information, in the reverse order as Alice encrypt the information. He can then recover then binary form of information, and finally gets what Alice sent him.

The scheme of the method is illustrated in the following figure.

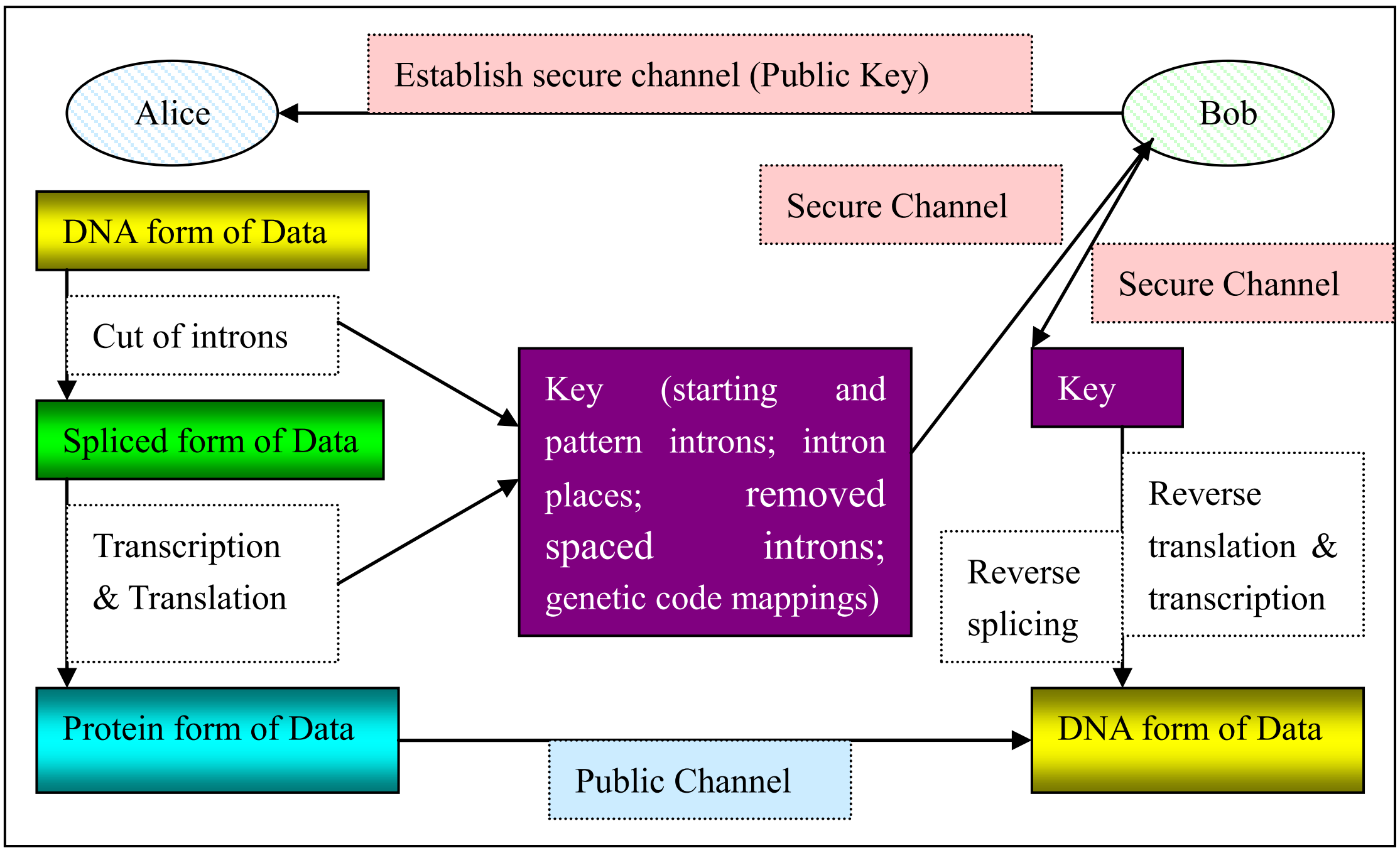

Figure 3: Basic scheme of the pseudo DNA cryptography method.

Generally, such a scheme can be formulated as below (M and C are for plaintext and cipher text, respectively)

- Encryption key E1 = (the starting and pattern codes of the introns, the places of the introns)
- Encryprion key E2 = (the codon-amino acids mapping)
- Decryption key D1 = E2
- Decryption key D2 = E1
- C' = E1(M), C = E2(C')
- M' = D1(C), M = D2(M')

The scheme is principally a symmetric key algorithm, except that the sender initially has only part of the keys, and he generates the rest part of the keys. It is obvious that such a scheme is essentially a 2 step substitution process, though they are substitutions in a general sense (not letter-by-letter substitution).

### Benefits and weaknesses of the method

The pseudo DNA cryptography method have many benefits, some of them are listed below:

1. Easy to encrypt the information by the sender, the sender does not need much information to encode. In other words, sender initially has part of the keys, and he generates other parts of the keys.
2. Little information needed to be communicated through a secure channel (only the keys). The key size is proportional to the size of the plaintext with a small ratio.

3. Protein form of information can be transferred through public channel, and the size of the protein form of information is generally smaller than that of the original information.
4. One-time pad can be possibly used as the key giving enough storage, since almost one key for one piece of information.

There are also quit some defects of the method, and We have listed some of which that WEfound not satisfactory here.

1. since the sender can not control the places of the introns, there is partial information available, which indeed includes pieces of plaintext. Multiple rounds can be used to relief this problem.
2. Analysis (such as differential analysis) may break the partial information of the cipher text.. A solution is that important texts are placed in intron (by selecting appropriate intron start and end patterns), and cipher text is also encrypted by other cryptography methods. By this means, the pseudo DNA cryptography method can be regarded as an enforcement of the traditional cryptography methods.
3. There may be bad intron starts and ends to cut introns. A solution is for sender to prepare many starting and ending codes of introns, and select a pair which can result in an appropriate cut off.
4. The more complex the keys, the more complex the decryption process.

### Method analysis

Suppose the DNA form of information D have the length $n$. There are $k$ introns, and the introns have the average length of $m$. Then the mRNA form of information D' have the length $n\text{-}k*m$. Since one codon (consists of 3 nucleic acids) generally can be translated into one amino acid, the protein form of information D'' have the length of $(n\text{-}k*m)/3$.

For the sender Alice, she only needs to scan the DNA form of information once, and cut off the introns to get the mRNA form of data, the time is in $O(n)$. After this, she can perform the translation process according to the genetic code table to generate the protein form of the data, also linear time. Therefore, Alice's encryption process have the time complexity of $O(n)$.

For the receiver Bob, he can use the keys that Alice sent him to perform the reverse operation. Since all of the genetic code mappings and positions of the introns are known, he can also perform the decryption process in $O(n)$ time (just two scans of the protein and mRNA form of information).

If Eve, and eavesdropper, can listen to Alice and Bob's communication, and he also gets the protein form of information, lets see how hard he have to try to break the encrypted information.

If Eve tries by brute force attack, then it is a very hard computational problem for him. In the protein to mRNA step, since there are 61 coding codons, and only 20 amino acids, in average there are 3 codons to be mapped onto the same amino acid, so that the different number of mRNA for the protein form of data is about $3^{(n\text{-}k*m)/3}$. In the reverse transcription process, if the continuous intron scheme is used, then there are $2^{n\text{-}k*m}$ possible combinations of places to insert introns, and even the introns are known, the time complexity of recovering the DNA form of data form mRNA form of data is $O(k*2^{n-k*m})$. If spaced intron scheme is used, the time complexity can be $O(2^k x 2^{n-k*m})$. So the total time complexity can be at least $O(3^{(n-k*m)/3} x 2^{n-k*m})$. If by careful control, $k*m$ is around $0.35n$ (which is $log_2 3/ (log_2 3+3)*n$), then the complexity above is $O(2^n)$. The smaller the $k*m$, such time complexity can be increased, but more partial information would be revealed, which makes it venerable to other attacks. Since n is generally of the 1M bits scale, such a brute force attack method is a disaster (unpractical) for Eve.

However, the method is not so strong against chosen plaintext attacks. Knowing the mechanisms of the method, and no knowledge about the keys (starting and pattern introns; intron places; removed spaced introns; genetic code mappings); the attacker Eve can get partial or full knowledge about the keys with high probability, if he is given enough plaintexts and computation. One way he can break the keys is a technique similar to that of the differential attack. The attacker can choose 2 plaintexts at a time, and these 2 plaintexts are different in only 1 or 2 places. The attacker can then encipher the 2 plaintexts by the pseudo DNA encryption method, and get the respective 2 cipher texts. By comparing and analyzing the cipher texts, the attacker can get information about some of the starting and pattern introns, as well as intron places and removed spaced introns. The comparison of different but similar pairs can reveal more information about these information, and some of the genetic code mapping information. Therefore, the more pairs the attacker used, the higher probability that he will recover the keys.

Suppose the DNA form of information D have the length $n$. There are $k$ introns, and the introns have the average length of $m$. Then the differential attacks can break the keys in relatively short time with enough plaintexts. In the optimal cases, the attacker can "predict" the intron places, he can then break one intron at a time; also he can recover the genetic code mapping using statistical analysis (if there are such statistical properties). In this case, the attacker can chose $2^{2m-1}$plaintexts for each intron test, and recover that intron with very high probability. Thus, the attacker will choose $k*2^{2m-1}$ plaintexts in total to break the keys. As regard to time, he needs $k*2^{2m-1}*m$ time to recover the introns, and linear or quadratic time to recover the genetic code mappings. Therefore, the attacker can in optimal cases break the keys in $O(k*2^{2m-1}*m)$ time. This is drastically smaller than that of the $O(2^n)$ time as brute force attacker have to use, especially in cases when in general $n>>m$. Though some the optimal assumptions (such as the statistical analysis and known intron places) are generally not satisfied, It is obvious that the differential attacks can recover the keys information with high probability in relatively short time and enough plaintexts. The use of multiple rounds, or more complicated splicing and mapping rules, may make the method stronger against these plaintext attacks.

The side channel cryptanalysis is also suitable for the pseudo DNA cryptography method, as the un-optimized operation time (or energy) is highly related with the content of the plaintext and key features (starting and pattern introns; intron places; removed spaced introns). The introduction of noises in the implementations can protect the method against some of such attacks.

Such a scheme is essentially a 2 step substitution process. Mainly substitution (and little transposition) is used in such a scheme. So that theoretically [Shannon, 1949], mostly confusion, and little diffusion of the plaintext can be achieved. This makes the scheme not so strong, and this situation can be partly relieved by the means of multiple rounds. A more practical application is for it to be used as an enhancement of the other cryptography methods, since it can greatly improve the security, especially against brute force attacks. Also, when multiple rounds of the method are used, it is perfectly sure that more transpositions would be introduced, and the security will be enhanced greatly.

Despite these analyses, We have to admit that some other critical problems are already found about this method, and We have listed some of them in the *Benefits and weaknesses of the method* part.

## 4. Experiments and Results

We have implemented the pseudo DNA cryptography method that We have explained. The experiments show this method to be efficient; and its power against certain attacks are shown by the theoretical analysis.

The programs are designed for sender and receiver. Generally, the programs for receiver perform the reverse processes as those processes the sender's program do.

On the sender side, the sender first needs his initial key, which includes the introns starting and pattern codes. These codes can be generated by a generator program, or by the user himself. The user also has the plaintext to encipher. He first translates the plaintext into DNA form of information using an information conversion program. Then he can simulate the splicing, transcription and translation process of central dogma with the respective program. In these processes, the necessary padding are also implemented for compatibility reasons. Through these steps, the starting and pattern codes of the introns, the places of the introns, the removed spaced introns, and the codon-amino acids mapping of the protein are added into the key file, and the enciphered information is also created. These two files can be then transferred to receiver through different channels (enciphered file trough public channel, and key file through secure channel).

On the receiver side, he receives the enciphered information and key file from different channels, and then uses the key information in the key file to decipher the enciphered information. He first performs the reverse translation, reverse transcription and reverse splicing process using the respective program, and the information stored in the key file. After these processes, he can get the DNA form of information, and he can then recover the plaintext using the recovery program. By these means, the receiver finally gets what the sender intended to tell him.

The implementation diagram of this cryptography method is illustrated in the following figure.

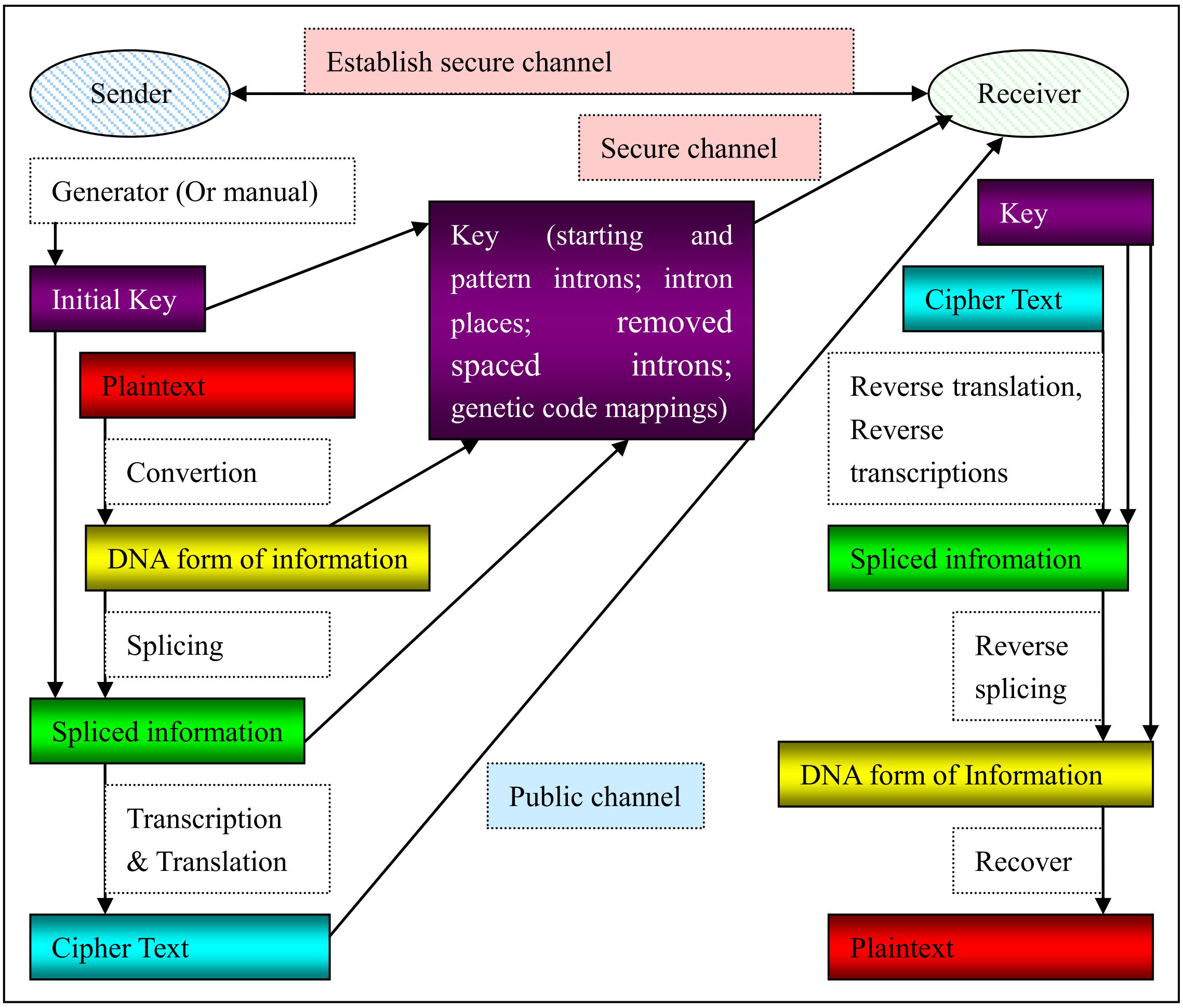

Figure 4: The implementation diagram of the pseudo DNA cryptography method.

Since We have stated in the *Benefits and weaknesses of the method* part that the existence of partial information makes the method not so strong by its own; expect for some attacks such as brute force attacks, the other attacks can break the current cipher text relatively easily. Therefore, WEwill only examine the power of the program theoretically. In the experiment part, We have focused on the efficiency and reliability of the program, and We have also examined the performance of the program as regard to the efficiency in storage and transmission. The results show that the program of the method is efficient and robust in computation, as well as in storage and transmission.

The experiments are performed under Linux environment, and many Linux console tools are used (s.a. time). The programs are written by Perl for clearness and efficiency, and the programs can be executed under UNIX compatible systems (without any configurations), as well as Windows systems (provided that Perl is installed in Widows). The test data are randomly selected from varies sources from the Internet, and they are stored on local machine for process.

The plaintexts that We have chosen are highly divert. They include short and long text, purely alphabetical text and text combining alphabets and many other characters. The objective is to test the performance of the program on different kinds of information.

As regard to the difference of the lengths of the text, We have selected 4 plaintexts with increasing size. The original plaintext size, the resulting cipher text size and the key size are examined; together with the encryption and decryption time. These features are the key features to examine the efficiency of the program in computation, storage and transmission.

Each of the plaintext has the length 10 times that of the former one, starting from 10. For each plaintext, We have also calculated the number of bits needed to store the plaintext in ASCII format, which is 8 times that of the length of the plaintext. For the cipher texts, We have listed the length of the cipher text, and the relative bits are also calculated. Since one amino acid can be represented by 3-letter sequence, and the 64 amino acids can be distinguished by 6 bits; only 2 bits are needed for each letter in the cipher text. Thus, the number of bits needed for each of the cipher text is 2 times that of the length of the cipher text. For the size of the key, the numbers listed include redundancies. The redundancies are those tags and separators that tell the receiver the corresponding information. The actual key information is dependent on the size of the starting and pattern codes of the introns; but generally, its size is roughly less than half that of the size including redundancy.

The encryption time and decryption time are also listed for each of the dataset. Under Linux system, each encryption and decryption process is performed 5 times, and the average system time is obtained and listed. This makes the evaluation of the time fair.

The results are shown in the following table.

| **Dataset** | **Length of Plaintext (Bits)** | **Length of Cipher Text (Bits)** | **Size of Key (with redundancy)** | **Encryption Time (ms)** | **Decryption Time (ms)** |
|---|---|---|---|---|---|
| Test1 | 10 (80) | 42 (84) | 96 | 244 | 423 |
| Test2 | 100 (800) | 312 (624) | 662 | 244 | 426 |
| Test3 | 1000 (8000) | 3303 (6606) | 4906 | 290 | 462 |
| Test4 | 10000 (80000) | 33321 (66642) | 47502 | 1285 | 1385 |

Table 1: The performance of application with plaintexts of different lengths.

The performance of the application can also be explained by the following plot.

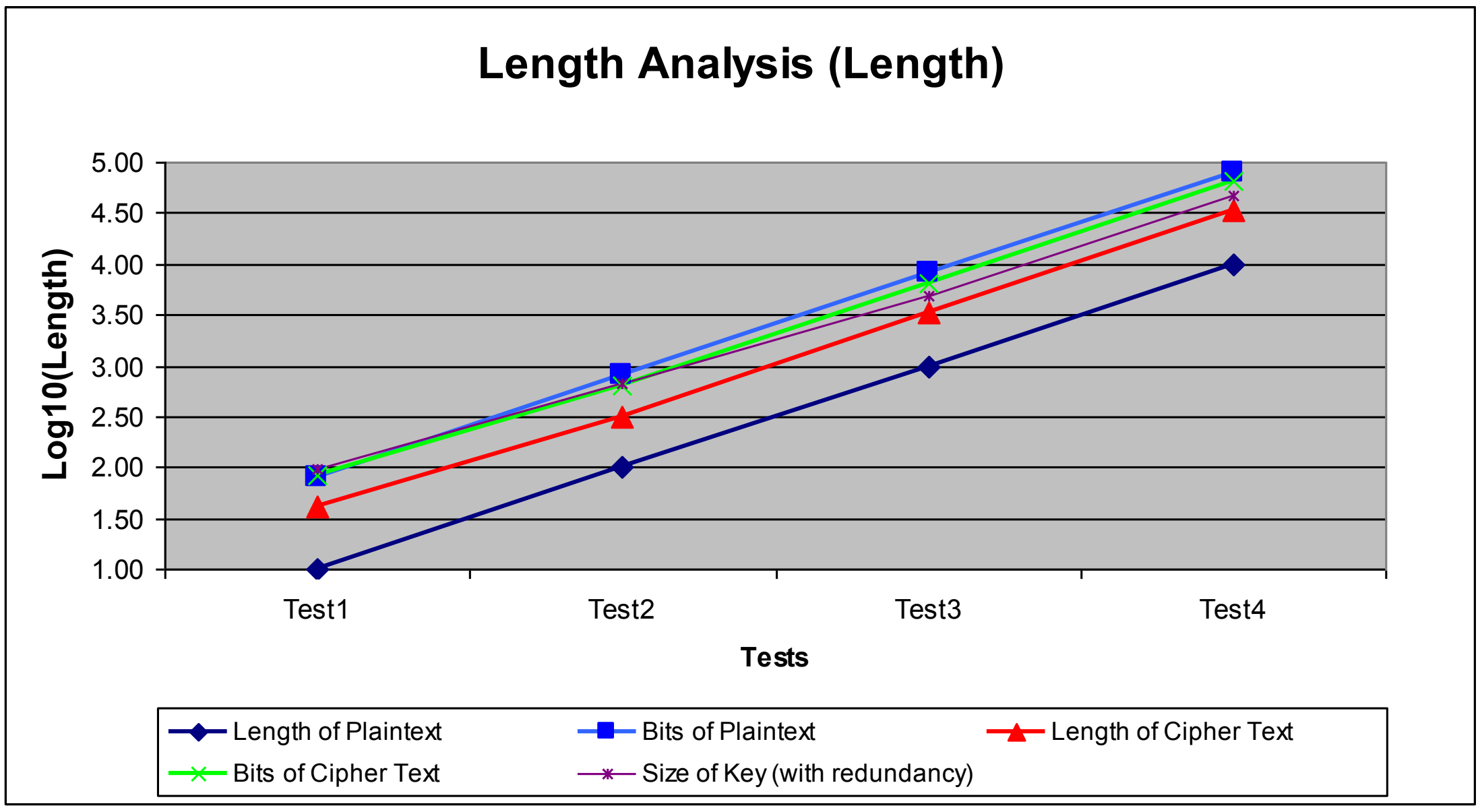

(A). Analysis of the length of plaintext and cipher text.

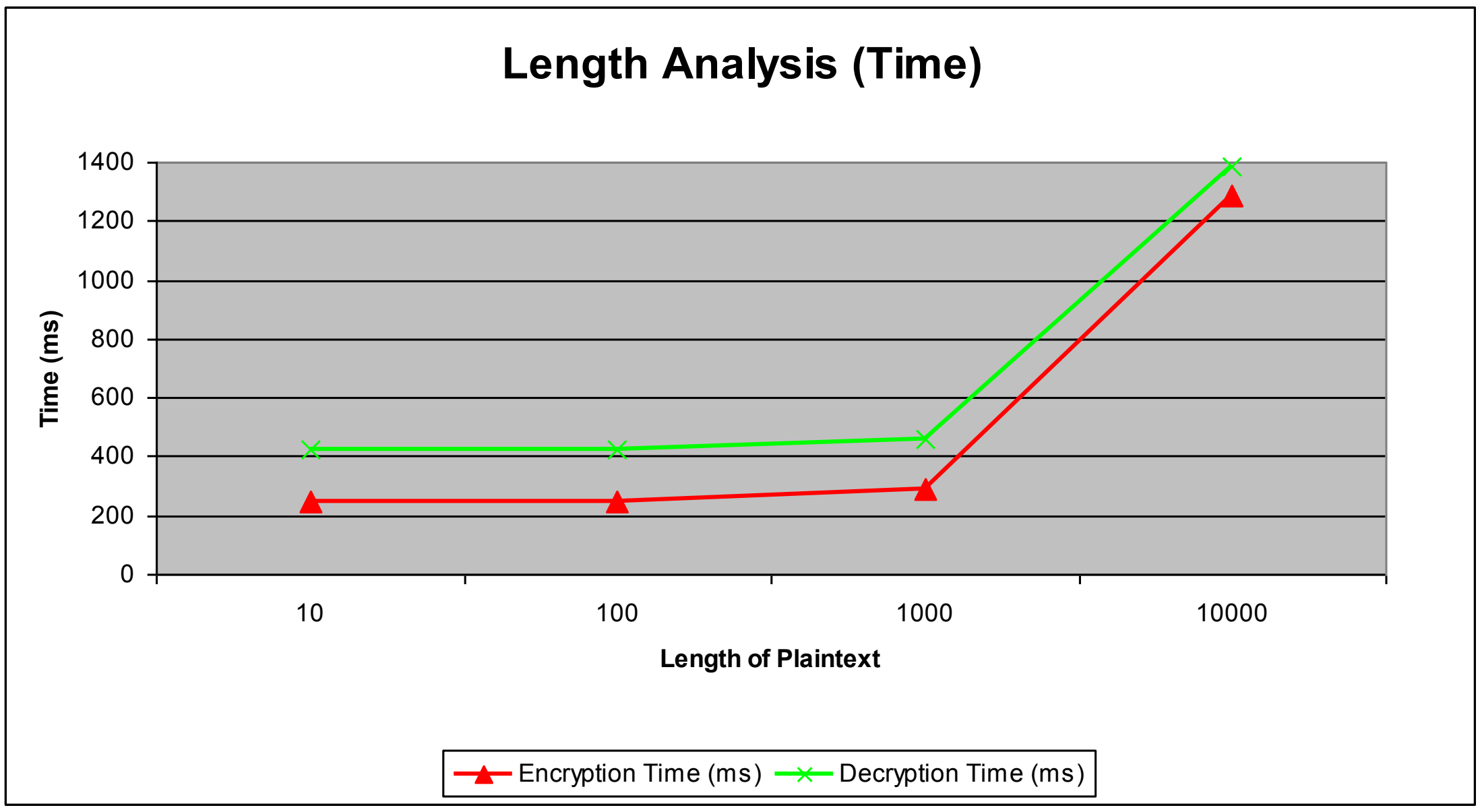

(B). Analysis of the time of enciphers and deciphers.

Figure 5: The performance of application with plaintexts of different lengths.

As shown in the table and plots. The cipher texts lengths are proportional to that of the corresponding plaintexts lengths. Indeed, when the actual bits are considered, the storage needed to store the cipher text is roughly ¾ of the storage needed to store the plaintext. This shows that the method is efficient in storage, since the cipher text need less storage space than that of the plaintext. Another observation is that the size of the key, even with high redundancy level, has the size increase as the size of the plaintext increase. Generally, the size of the redundant key is round 5.0 times (ranging from 10.0 to 4.5 times, and approaching 5.0 when the length of plaintext increase) that of the size of the plaintext. If redundancies are removed, and proper representations of the information in the key file are chosen (for example, 2 bits to represent nucleic acids, rather than ASCII representation), the key size would then be reduced sharply to a fraction of the size of the plaintext. Because of these, the cipher text can be transferred through public channel fast (even faster than transmitting the plaintext), and the small-sized key can also be transmitted fast through the secure channel. Thus, the method is efficient in both storage and transmission.

The encryption and decryption time show that the program is efficient in computation. The relative table and plots show that the encryption and decryption time for the different length of the plaintext increase slower than the changes in the length of the plaintext. This reveals that the processing time can be very fast even for relatively very long plaintext. Therefore, the efficiency in computation is obvious. This indicates the potential applicability of the multiple rounds methods.

As to the different content of the text, we have chosen 5 plaintexts with different contents. The objective is to check the robustness of the program.

The plaintext datasets are selected so that they are very representative. One plaintext contains only alphabetical and digital characters, one only contains non-alphabetical and non-digital characters, and other 3 contains the combination of characters with several unusual cases. By this means, the examination on robustness can be performed with high confidence.

The results are shown in the following table.

| **Dataset** | **Description** | **Number of Different Characters** | **Recovery of Plaintext** |
|---|---|---|---|
| Test1 | Only alphabetical and digital characters | 52 | Yes |
| Test2 | Only non-alphabetical and non-digital characters | 100 | Yes |
| Test3 | Combination of characters | 75 | Yes |
| Test4 | Combination of characters | 125 | Yes |
| Test5 | Combination of characters | 150 | Yes |

Table 2: The performance of application with plaintexts of different contents.

The experiments on plaintext with different contents are performed to test the robustness of the program, and the results show that the program can handle almost all of the characters and their combinations with high accuracy and efficiency, and this proves that the program is very robust.

The experiments that We have done are currently for explanations of the method. Therefore, the key data have redundancies; the DNA, RNA and protein form of information (including cipher texts and keys) are not in their binary form. The more elegant and realistic program using binary representation of the information, together with better organization of the key files, will surely enhance the effectiveness and robustness of such a cryptography method. These stuffs, together with some extensions and variations of the method, are under careful investigations now, and they will be ready later.

In Appendix we have listed some important source codes, some sample results, and the standard key file format with examples.

## 5. Conclusions and Future Works

In this project, we have proposed pseudo DNA cryptography method, a new cryptography method based on the central dogma of molecular biology. The method simulates the transcription and translation process of the central dogma; it also adds some artificial features to make the resulting cipher texts difficult to break. The theoretical analysis shows that this method is powerful against certain attacks, especially against brute force attacks. The experiments not only show the power of such a method, but also reveal that this method is very efficient in computation as well as in storage and transmission, and it is very robust.

However, since it is only in its primitive stage, this method has some defects, and it needs quite some improvements. The most serious problem is that partial information exists, so that attackers can infer the whole information with only some effort. This is the case even the spaced introns are introduced. The probable reason is that the confusion of the plaintext is not very complete, so that the relationship between statistics of cipher text and statistics of plaintext is not as complex. one solution is to do multiple rounds of encryption on plaintext to get the cipher text. Other problems are related with the degree of diffusion and confusion of the method; and complexity of decipher. This cryptography system also has several implementation defects now, which can be overcome by better representations of the information and organization of the data in the files. In the future, the theoretical analysis on these problems will be performed, and related experiments will be performed to examine the performance.

As this method is still primitive now, it has a great scope to have some extensions, as well as some variations. Many of these extensions and variations come from the modification of the central dogma principle. And these changes can always result in more secure methods against attacks.

The variations based on this pseudo DNA cryptography method are interesting. One such variation is to artificially modify the codon-amino acids mapping, so that it is more complicated, saying, one protein mapping to more than 3 amino acids. By this means, the combination of mappings can be increased potentially, and make the brute force attacks more difficult. Another variation is to make the splicing rule more flexible, and this can enhance the confusion effects. The effect of such variations, together with some others, will be scrutinized in the future.

As to extensions of the method, to use multiple rounds of the method is mentioned on the above, trying to solve the partial information problem. This is a general idea in cryptography, and the experiments about it will be performed later on.

The pseudo DNA cryptography method that we have proposed is not constraint to the encryption and decryption area, but also such areas like massage authentication. Indeed, the encryption process of the method can be very powerful and more effective if it is used as a hash function for generating message authentication code. By this means, only the initial key is needed, and the encryption process can be fast. If multiple rounds are used, the security may also be satisfactory. The problem is that the length of the resulting MAC is hard to control, and one of the solutions to this problem is to do multiple round of the process, together with padding to get the MAC with fix length. If using real DNA data, the steganography can be implemented, especially for image steganography. The use of this method to do message authentication, as well as its other usages, will also be examined in the future.

This method based on the central dogma has its nature advantage in software and hardware applications. Since only 4 nucleotides are involved in encipher and decipher process of the plaintext, the program can be turned to be a very efficient program based on primitive codes in C or assembly language. The hardware implementation can be easy based on these simple principles, and it can also be very efficient. With today's advanced hardware technology, the multiple rounds can be easily implemented, and the security can be relatively reasonable. These benefits show that this method is very suitable for software and hardware implementation. The further analysis and experiments on these factors is very interesting topics in the future works.

## Appendix

### Appendix A: Source File Names and Brief Descriptions

| **Sender** | | | |
|---|---|---|---|
| **File Name** | **Description** | **Input** | **Output** |
| Key_generation.pl | Generation of initial keys | Seed to generate random numbers,<br>Random number selector,<br>Length of the starting code,<br>Length of the pattern code | Initial Key (Starting codes,<br>Pattern codes) |
| Info_convertion.pl | Converting input information to DNA form | Plaintext | DNA form of information |
| Splicing.pl | Splicing process | Initial Key,<br>DNA form of information | Spliced information,<br>Key (partial) |
| Translation.pl | Transcription and Translation process | Spliced information,<br>Key (partial) | Cipher text,<br>Key (complete) |
| DNA_encryption.pl | The whole DNA encryption process | Plaintext file | DNA form of information file, spliced information file, enciphered information file |
| **Receiver** | | | |
| **File Name** | **Description** | **Input** | **Output** |
| Reverse_translation.pl | Reverse Translation and reverse Transcription process | Cipher text,<br>Key (complete) | Spliced information |
| Reverse_splicing.pl | Reverse splicing process | Spliced information,<br>Key (complete) | DNA form of information |
| Information_recovery.pl | Information recovery from DNA form | DNA form of information | Plaintext |
| DNA_decryption.pl | The whole DNA decryption process | Cipher text file | spliced information file, DNA form of information file, plaintext file |

**Appendix B: Some Important Source Codes**

**DNA_encryption.pl**

```
#Filename  : DNA_encryption.pl
#Author    : Ning Kang
#Version   : 1.0
#Date      : 2004/10/02
#Description  : DNA encryption process

$plaintext="@ARGV[0]";

if(@ARGV < 1){
    print "Usage perl DNA_encryption.pl <plaintext file>\n";
}
else{
    $dnatext=$plaintext."_dna";
    $splicedtext=$plaintext."_spliced";
    $ciphertext=$plaintext."_enciphered";

    system "perl Info_convertion.pl < $plaintext > $dnatext";
    system "perl Splicing.pl Init_key Key < $dnatext > $splicedtext";
    system "perl Translation.pl Key < $splicedtext > $ciphertext";
}
```

**DNA_decryption.pl**

```
#Filename  : DNA_decryption.pl
#Author    : Ning Kang
#Version   : 1.0
#Date      : 2004/10/02
#Description  : DNA decryption process

$ciphertext="@ARGV[0]";

if(@ARGV < 1){
    print "Usage perl DNA_decryption.pl <ciphertext file>\n";
}
else{
    $splicedtext=$ciphertext."_spliced";
    $dnatext=$ciphertext."_dna";
    $plaintext=$ciphertext."_deciphered";

    system "perl Reverse_translation.pl Key < $ciphertext >
```

```
$splicedtext";
    system "perl Reverse_splicing.pl Init_key Key < $splicedtext >
$dnatext";
    system "perl Information_recovery.pl < $dnatext > $plaintext";
}
```

**Appendix C: Sample results**

| **Sender** | | | |
|---|---|---|---|
| **File Name** | **Generator** | **Description** | **Content** |
| Init_key | Key_generation.pl | Initial key for sender | TAG<br>1010010010 |
| test | | Plaintext | We have a "secret"! |
| test_dna | Info_convertion.pl | DNA form of information | CAGCAGAACGGACGACCTCGCGCCAGAACGACAGAAAGAGCTATC<br>GCCCGATCTAGCGCCCTCAAGAGAGAC |
| test_spliced | Splicing.pl | Spliced information | CAGCAGAACGGACGACCTCGCGCCAGAACGACAGAAAGAGCTATC<br>GCCCGATCCCTAAGAGAC |
| test_enciphered | Translation.pl | Encrypted information after transcription and translation | ValSer...GlySerGlyAspSerSerPheCysArgSerGlyAla<br>ArgSerSerValLeuLeu |
| Key | Key_generation.pl,<br>Splicing.pl,<br>Translation.pl | Key to decipher the cipher text | TAG<br>1010010010<br>----------<br>splicing.position: 53<br>splicing.deleted: GCCCAG<br>----------<br>mapping: 10123105010001353105 5<br>---------- |
| **Receiver** | | | |
| **File Name** | **Generator** | **Description** | **Content** |
| test_enciphered_spliced | Reverse_translation.pl | Spliced information | CAGCAGAACGGACGACCTCGCGCCAGAACGACAGAAAGAGCTATC |
| | | | GCCCGATCCCTAAGAGAC |
| test_enciphered_dna | Reverse_splicing.pl | DNA form of information | CAGCAGAACGGACGACCTCGCGCCAGAACGACAGAAAGAGCTATC GCCCGATCTAGCGCCCTCAAGAGAGAC |
| test_enciphered_deciphered | Information_recovery.pl | Recovered plaintext | We have a "secret"! |

**Appendix D: Key File Format**

The standard key file generally includes the starting and pattern codes of the introns, the places of the introns, the removed spaced introns, and the codon-amino acids mapping of the protein. Below are contents in a standard key file.

```
TAG                                     (starting codes)
1010010010                              (pattern codes)
----------
splicing.position:   53                 (place of the intron)
splicing.deleted: GCCCAG                (removed spaced intron)
----------
mapping:  10123105010001353105 5        (codon-amino acids mapping)
----------
```

The starting and pattern codes of the introns are set by the initial key generator, or manually by users. Below are contents in a standard initial key file.

```
TAG                                     (starting codes)
1010010010                              (pattern codes)
```